\documentclass[epj]{svjour}
\usepackage{hyperref}
\usepackage{doi}
\usepackage{graphics}
\usepackage{epsfig}
\usepackage{soul}
\usepackage{color}
\usepackage[utf8]{inputenc}

\usepackage[sort&compress,numbers]{natbib}
\renewcommand{\abstractname}{\vspace{-0.1cm}}
\def\makeheadbox{\relax}

\titlerunning{Comment on ``How (not) to renormalize integral equations...''}

\begin{document}
\title{Comment on ``How (not) to renormalize integral equations with singular potentials in effective field theory'' }
\author{Manuel Pavon Valderrama\inst{1}%
}                     
%
%
\institute{School of Physics and Nuclear Energy Engineering, \\
International Research Center for Nuclei and Particles in the Cosmos and \\
Beijing Key Laboratory of Advanced Nuclear Materials and Physics, \\ 
Beihang University, Beijing 100191, China, \email{mpavon@buaa.edu.cn}
}
\date{Received: date / Revised version: date}
%
\abstract{[
  I critically discuss two of the potential inconsistencies
  pointed out in the recent manuscript by Epelbaum, Gasparyan, Gegelia
  and Mei{\ss}ner, published in Eur. Phys.J. A54, 186 (2018);
  the conclusion is that these inconsistencies do not happen.]
}
%
\maketitle

In a recent manuscript Epelbaum et al.~\cite{Epelbaum:2018zli} discuss
{\it non-perturbative renormalization} within
effective field theory (EFT).
The term refers mostly to renormalization as applied
in a purely non-perturbative context,
but tangentially includes
the mixture of perturbative and non-perturbative renormalization
advocated, for instance, by Nogga et al.~\cite{Nogga:2005hy}.
The idea is that {for really hard cut-offs} this type of
renormalization will lead to a series of (apparent) inconsistencies.
Two possible inconsistencies discussed in detail are:
(i) a possible mismatch between the $\hbar$ expansion of the scattering
amplitude and renormalization, (ii) an impossibility of
{\it non-perturbative renormalization} to deal with
repulsive singular interactions.
Here I review these {\it inconsistencies}.

First, I will discuss the $\hbar$ mismatch. It can be explained
as follows: we consider a scattering amplitude $T$
in a theory with a cutoff $\Lambda$.
The amplitude $T$ is generated from the iteration of
the leading order potential $V = C_0 + C_2 (p^2 + p'^2)$.
$T$ has a well-defined $\Lambda \to \infty$ limit, from which we
are tempted to conclude that it has been renormalized
in agreement with EFT principles.
But when we consider its $\hbar$ expansion, i.e. $T = T^{[0]} + \hbar\,T^{[1]}
+ \mathcal{O}(\hbar^2)$, a closer inspection reveals that $T^{[1]}$
contains a linear divergence.
That is, according to Ref.~\cite{Epelbaum:2018zli} the amplitude $T$ has
not been renormalized in agreement with the EFT principles
{(notice that this is controversial because $T^{[1]}$ is not observable,
  but let us concede that there is an inconsistency in what follows).}
The diagnosis of Ref.~\cite{Epelbaum:2018zli} is that the problem lies
in a misguided insistence in taking the $\Lambda \to \infty$ limit,
which is not necessary within the EFT framework and
can potentially lead to more harm than good.
Though the inconsistency {might indeed be there},
this diagnosis is incorrect
owing to a subtlety in the renormalization process that is regularly
ignored in cutoff regularization (because, except for formal
settings as this one, it is generally of no consequence).
This subtlety is the existence of two contributions to the EFT counterterms,
one that contains physical information and one that doesn't.
I refer to the second type of contribution as redundant counterterms (RC),
which are discussed in Ref.~\cite{Valderrama:2016koj}.
The RCs are there to cancel the residual cutoff dependence,
but can be ignored in practice when $\Lambda \to \infty$.
RCs are however regularly included in EFTs using
power divergence subtraction (PDS) regularization~\cite{Kaplan:1998we},
for instance.
In cutoff EFT a trivial way to remove the divergences in the $\hbar$ expansion
is to explicitly include the RCs, which for the $T$ matrix discussed
here take the form $V^{R} =  \sum_{n \geq 2} C^R_{2n} (p^{2n} + p'^{2n})$,
with {$C^R_{2n} = \sum_k \hbar^{k}\, {C^R_{2n}}^{[k]}$}.
{The number of RCs is infinite, but this is inconsequential because
  they are not observable and carry no physical information;
  instead, they are an analysis tool.}
Concrete calculations for this redundant potential are easy to perform with PDS,
where explicit solutions exist for the $C_{2n}^R$ and where it can be explicitly
shown that no positive power of $\Lambda_{\rm PDS}$ (i.e. the PDS {\it cutoff})
appear in the $\hbar$ expansion of $T$.
With a sharp cutoff in momentum space, calculations are more involved and
cannot be solved in closed form, yet it is trivial to check
that ${C^R_{4}}^{[1]}$ removes the $\hbar$ inconsistency.

\begin{figure}
  \begin{center}
    \epsfig{figure=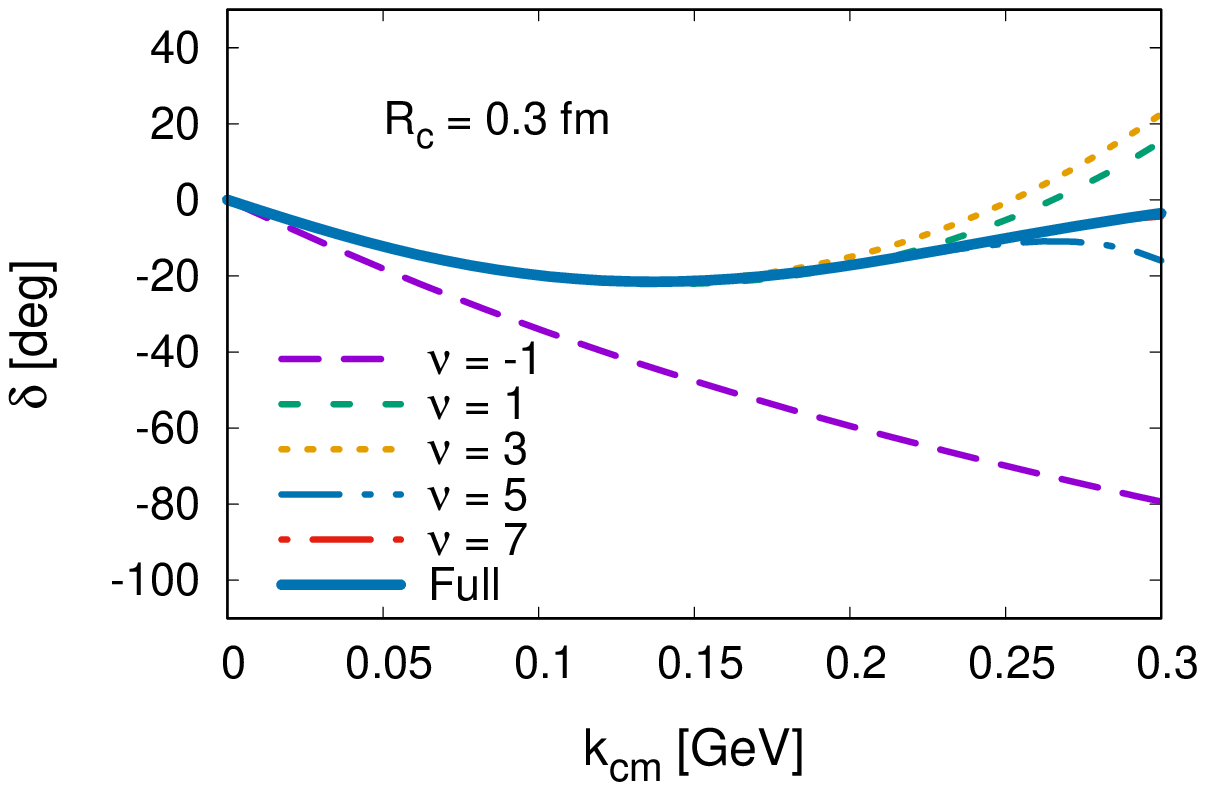,
      height=5.0cm, width=8.0cm}
    \epsfig{figure=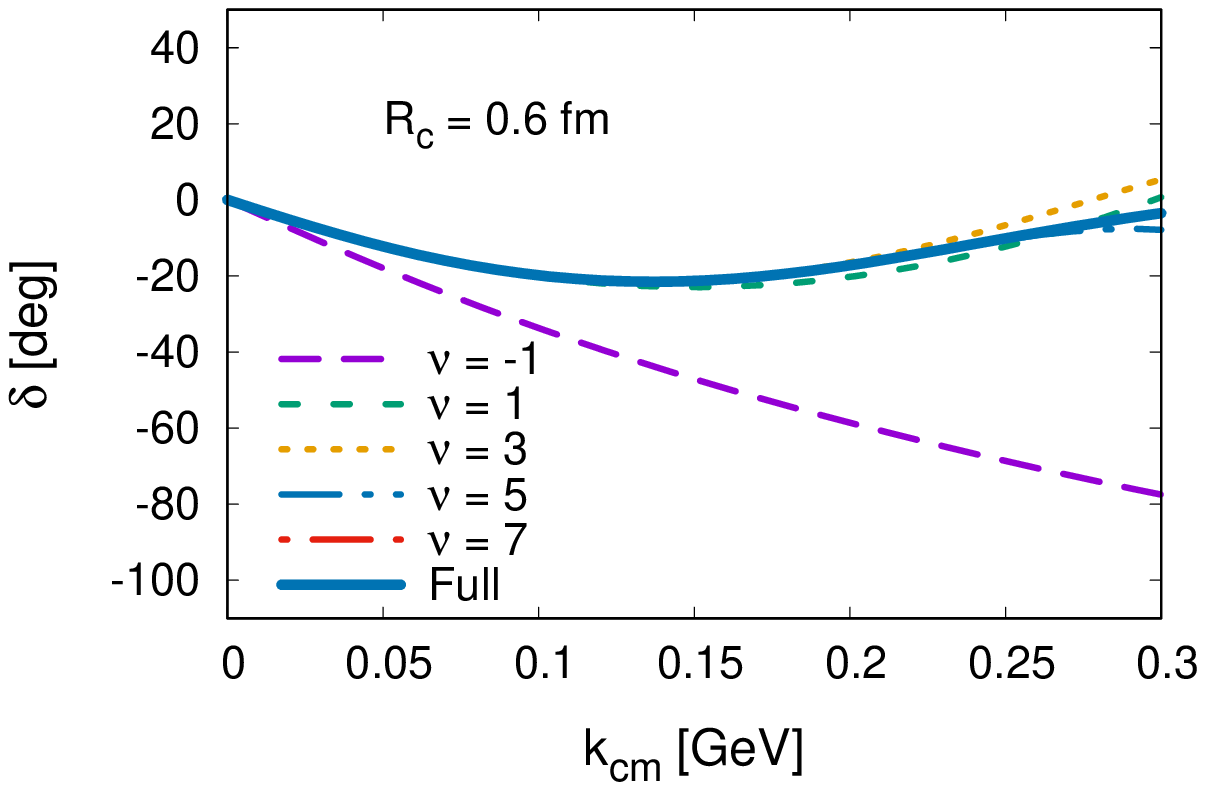,
      height=5.0cm, width=8.0cm}
    \epsfig{figure=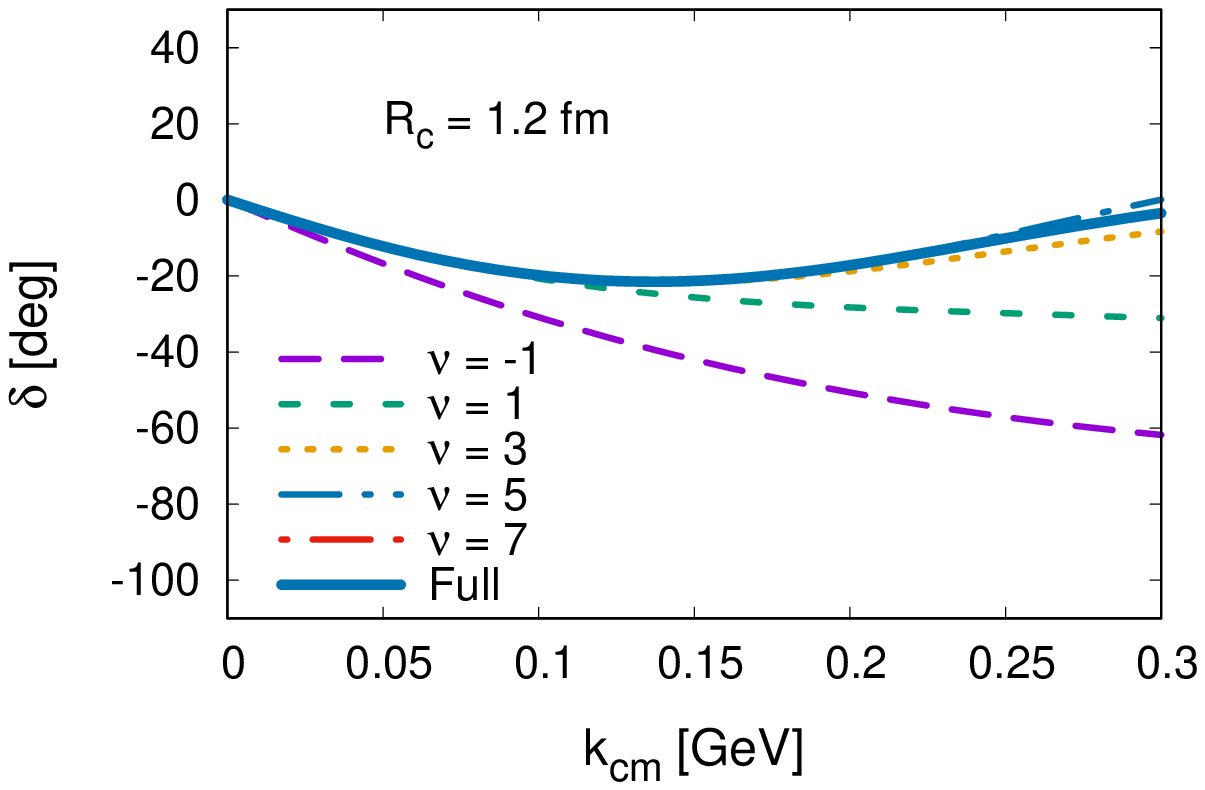,
      height=5.0cm, width=8.0cm}
    \epsfig{figure=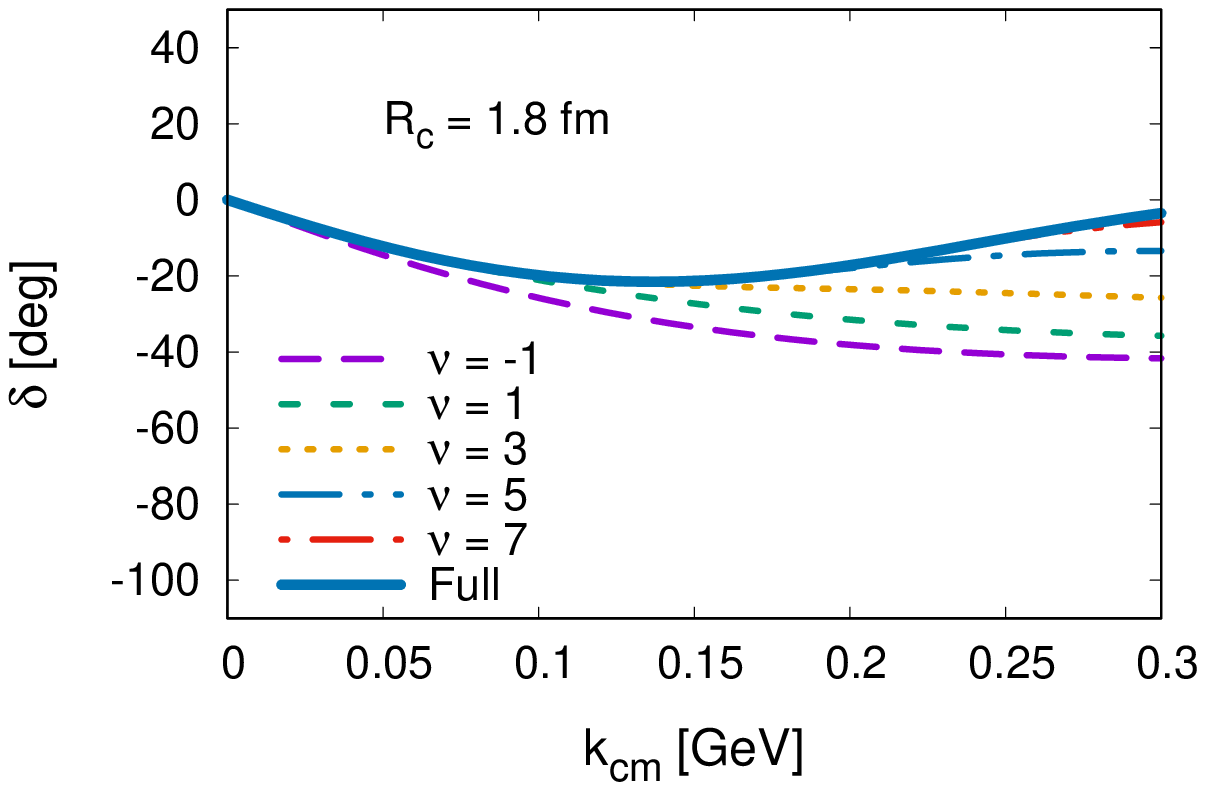,
      height=5.0cm, width=8.0cm}
  \end{center}
  \caption{Phase shifts for the toy model of Ref.~\cite{Epelbaum:2018zli}
    computed within a suitable EFT which takes into account
    the repulsive singular nature of the leading order potential.
    In the power counting we use for the toy model,
    $C_0\,\delta(\vec{r})$ enters at order $Q^{1}$,
    $C_2\,\nabla^2 \delta(\vec{r})$ at order $Q^{3}$,
    $C_4\,\nabla^4 \delta(\vec{r})$ at order $Q^{5}$ and
    $C_6\,\nabla^6 \delta(\vec{r})$ at order $Q^{7}$.
    The $Q^7$ calculation usually falls on top of the full one.
    The contact interactions are iterated according to the counting, e.g.
    at order $Q^3$, $C_0$ is iterated once and $C_2$ enters
    at tree level.
  }
\label{fig:EFT-repulsive}       
\end{figure}

Second, I will discuss the {\it non-perturbative renormalization} of
the interesting toy model of Epelbaum et al.~\cite{Epelbaum:2018zli},
which contains a repulsive singular interaction at leading order.
The potential in this toy model contains a long- and short-range piece,
$V(r) = V_L + V_S$, with $V_L$ singular and repulsive and $V_S$ attractive.
Despite the singular nature of $V_L$, the full potential $V$ is regular,
where the details of the toy model can be consulted
in Ref.~\cite{Epelbaum:2018zli}.
As shown by explicit calculations $V_L$ is not renormalizable
if treated purely non-perturbatively in combination
with a contact-range interaction~\cite{Epelbaum:2018zli}
(in agreement with Ref.~\cite{PavonValderrama:2005uj}).
Here I stress that the inconsistency is not in the renormalization process,
but in the power counting employed: EFT requires to iterate according
to power counting, but repulsive singular interaction likely entail
a demotion of the contact interactions, not a promotion, 
see the discussion below (or see Ref.~\cite{Birse:2005um}
for a different opinion).
The authors of Ref.~\cite{Epelbaum:2018zli} put into question that
a mixture of non-perturbative and perturbative methods could
reproduce the fundamental theory.

This viewpoint, though sensible, is premature:
here I present calculations for this toy model within a
mixture of perturbative and non-perturbative renormalization.
For that I simply include the long range potential $V_L$ as the leading
order of the calculation and include contact interactions
according to their power counting, which I determine
by adapting the ideas of Ref.~\cite{Valderrama:2014vra}.
The outcome is that $V_L$ will be counted as $Q^{-1}$
(to justify its iteration), while the contact-range couplings
will be demoted by one order with respect to naive dimensional analysis,
i.e. $C_{2n}(p^{2n} + p'^{2n})$ enters at order $Q^{2n+1}$.
Concrete calculations of the phase shifts are shown
in Fig.~\ref{fig:EFT-repulsive},
where we can see that the perturbative expansion indeed converges well.
We regularize the long-range potential with a Gaussian regulator
in coordinate space
\begin{eqnarray}
  V_L(r) \to V_L(r; R_c) = V_L(r)\,(1 - e^{-(r/R_c)^2})\, ,
\end{eqnarray}
while for the contact-range potential we regularize the Dirac-delta as 
\begin{eqnarray}
  \delta(\vec{r}) \to \delta(r; r_c) = \frac{1}{\pi^{3/2} R_c^3}\,e^{-(r/R_c)^2}
  \, ,
\end{eqnarray}
plus similar expressions for its derivatives, where a local
contact-range potential is used: $C_{2n} q^{2n}$.
The regularization is slightly different than in the original manuscript,
but it is certainly simpler and nonetheless equivalent.
{The details of the calculation are analogous to those of
  Ref.}~\cite{Valderrama:2011mv}, {but extended to higher orders.}
{The $C_{2n}$ couplings are determined by fitting to the toy model
phase shifts in the $20-80\,{\rm MeV}$ ($80-200\,{\rm MeV}$) range
for $\nu = 1,3$ ($\nu = 5,7$).}
Calculations are shown for the cutoffs $R_c = 0.3, 0.6, 1.2$ and
$1.8\,{\rm fm}$ up to order $Q^7$ ($\rm N^8LO$) in the EFT expansion.
The conclusion is that the standard EFT approach of Ref.~\cite{Nogga:2005hy}
is perfectly able to describe the physics of
the toy model of Epelbaum et al.~\cite{Epelbaum:2018zli}.
In addition it improves over the proposal of Ref.~\cite{Epelbaum:2018zli}
(namely, a purely non-perturbative approach
with a judiciously chosen cutoff),
in the sense that there are no  {strong} restrictions
on the cutoff
(besides the numerical ones, {$R_c \geq 0.3\,{\rm fm}$ in this case}),
which can be taken harder than the breakdown scale if one wishes to.
{Notice that even though the existence of the $R_c \to 0$ limit
  has not been proven, this is not a necessary condition
  for the present approach to be useful.}

\begin{acknowledgement}
This work is partly supported by the National Natural Science Foundation
of China under Grants No.11522539, No.11735003, the Fundamental
Research Funds for the Central Universities and the Thousand
Talents Plan for Young Professionals.
\end{acknowledgement}

%
%

%

\end{document}